\documentclass[prl,aps,twocolumn]{revtex4} 
\usepackage[dvipdfmx]{graphicx}
\usepackage{amsmath,amssymb,amsfonts}

\def\be{\begin{equation}}
\def\ee{\end{equation}}
\def\l{\left}
\def\r{\right}

\newcommand{\vect}[1]{\mbox{\boldmath $ #1 $}}

\begin{document}

\title{Multiple Half-Quantum Vortices in Rotating Superfluid $^3$He}
 
\author{Mikio Nakahara$^{1,2}$ and Tetsuo Ohmi$^1$}
  
\affiliation{$^1$Research Center for Quantum Computing,
Interdisciplinary Graduate School of Science and Engineering,
Kinki University, Higashi-Osaka, 577-8502, Japan\\
$^2$Department of Physics,
Kinki University, Higashi-Osaka, 577-8502, Japan
}

\begin{abstract}
Half-quantum votices and ordinary vortices in a rotating thin film
superfluid $^3$He under a strong magnetic field are considered.
It is shown that $2n+1$ half-quantum vortices interpolates 
between $n$ singular vortices and $n+1$ singular vortices
as the angular velocity is changed. The phase
diagram of the vortex configurations in the 
angular velocity-magnetic field space is obtained
for a paramagnon parameter $\delta=0.05$.
\end{abstract}

\maketitle

Superfluid $^3$He exhibits extremely exotic and interesting
properties due to its complex order parameter \cite{vw,vol1,vol2},
which attracted much attention not only of condensed matter physicists
but also of particle theorists and gravitational physicists.
One of the manifestations of such exotic properties is a vortex having a 
half amount of
vortex quantum called a helf quantum vortex, abbreviate as HQV
hereafter, whose existence was predicted first by
Volovik and Mineev in 1976 \cite{vm,sv}. 
A HQV is also expected to be present in BEC of alkali atoms
\cite{bec2,bec3,bec4} and spin-triplet superconductors \cite{sc1,sc2,sc3}, 
among other physical systems.
In spite of extensive theoretical
\cite{th1,th2,th3} and
experimental \cite{ex1,ex2} research on HQV in superfluid $^3$He since then,
its existence is yet to be experimentally confirmed.

Recently, we investigated a rotating superfluid $^3$He in a slab geometry
under a strong magnetic field \cite{on}, in which 
we have shown that a HQV is energetically stable compared
to a singular vortex (SV) in the A$_2$-phase side (i.e., lower temperature
side) in the vicinity of the A$_1$-A$_2$ phase boundary. 
In this part of the phase diagram, a HQV will nucleate first
as the angular velocity of the rotation is increased from
zero. Let us summarize our results obtained in \cite{on} to establish
notations and convention. 
Consider a rotating thin film of superfluid $^3$He in a cylindrical 
slab geometry under a strong magentic field $H$. In the presence of a
magnetic field, the superfluid
has different populations between the spin up-up $(+)$ condensate
and the spin down-down $(-)$ condensate, where the spin direction
is measured with respect to the magnetic field.
This phase is called the A$_2$ phase.
The angluar velocity $\vect{\Omega}$ is parallel to
the $z$-axis and the film is perpendicular to the $z$-axis. 
The magnetic field is taken parallel to the rotation axis.
The thickness and the radius of the film are denoted by $d$ and $R$, 
respectively, where $d$ must be less than the dipole 
coherence length so that the $\hat{\vect{d}}$-vector stays in the
$xy$-plane throughout the condensate. We use the Ginzburg-Landau free energy
\cite{vw}, in which the correction of the fourth order coefficients 
$\beta_i$ of the bulk
free energy by the paramagnon parameter $\delta$ is taken into account,
to find the most stable vortex configuration for given parameters
$\delta$, $\Omega$
and $H$. 

Instead of expanding the order parameter in terms of the standard
Cartesian base $\{\vect{e}_i \}=\{\vect{e}_x, \vect{e}_y, \vect{e}_z\}$, we expand it
in terms of $\{\vect{e}_{\mu}\}=\{\vect{e}_{\pm}, \vect{e}_0\}$ base
defined by
$$
\vect{e}_{\pm}= \mp \frac{1}{\sqrt{2}}(\vect{e}_x \pm i \vect{e}_y),
\ \vect{e}_0 = \vect{e}_z.
$$
The boundary condition $\hat{\vect{l}}=\pm \hat{\vect{z}}$ 
forces $A_{\nu \pm}$ have non-vanishing values in the bulk. 
Here the first subscript of $A$ is the spin index
while the second one is the orbital index. 
A strong magnetic field along the $z$-axis further forces the 
order parameter to have only four nonvanishing components $A_{\pm \pm}$
in the bulk. Let $t = 1-T/T_c$, $T_c$ being the critical temperature,
and $\alpha = \alpha' t$, where $\alpha$ is the coefficient of the second
order term of the bulk free energy and define the 
scaled magnetic field $h$ by $h=\eta H/\alpha'$, where
$\eta$ is a constant coupling strength between $H$ and the
condensate. It turns out to be convenient to further scale $h$ as
$\hat{h} = h/t$. The bulk order parameter is found by minimizing
the uniform Ginzburg-Landau free energy. We assume the vortex is embedded 
in a $\hat{\vect{l}}=+\hat{\vect{z}}$ texture, for concreteness, and the order
parameter has only nonvanishing components $A_{\pm  +}$ at $r \gg 1$,
where the length is scaled by the coherence length with
vanishing external magnetic field.
We parameterize the components as $A_{\mu\nu} = C_{\mu \nu}(r)
e^{i n_{\mu\nu} \phi}$ assuming the cylindrical symmetry, where $\phi$ is
the azimuthal angle in the $xy$-plane and $n_{\mu\nu} \in \mathbb{Z}$. 
It turns out that $n_{\mu\nu}$ satisfy the quantization condition 
$n_{\mu -}= n_{\mu +}+2$
due to the coupling between $A_{\mu +}$ and $A_{\mu -}$
through the gradient free energy \cite{on}.

When the HQV order parameter is expanded in $\{\vect{e}_{\nu}\}$,
it is found that the order parameter is a superposition of 
a $(+)$ condensate with no winding number and a $(-)$ condensate
with a unit winding number or the other way around. 
Such a HQV has a free energy
\be
F_{\rm HQV}^{(\pm)}= 2\pi \int_0^R r dr (F-F_0) = 4 \pi \l(A_{\pm +}^{(0)}
\r)^2 \l( \ln R+C_{\pm}\r), 
\ee
where $F_0$ is the bulk free energy without a vortex. 
Here $A_{\pm +}^{(0)}$ stands for
the amplitude of the bulk order parameter with orbital state $l_z =+ 1$
of the up-up ($+$) or the down-down ($-$) spin condensate, while $C_{\pm}$
is the vortex core energy of the $(\pm)$ condensate. 
The parameters $C_{\pm}$ are
obtained as functions of the paramagnon parameter $\delta$
(i.e., the pressure) and the external magnetic field numerically \cite{on}. 
The inequality $C_-
<C_+$ is always satisfied since the coherence lenghts $\xi_{\pm}$
of the condensates $(\pm)$ satisfy $\xi_+<\xi_-$, form which we
find a HQV carrying a vortex of unit winding number in the $(-)$ condensate
 has less energy compared to 
that with a vortex in the $(+)$ condensate. Let $L^{(-)}
= 4\pi(2m/\hbar) (A_{-+}^{(0)})^2 R^2$ be the angular momentum of the system.
By considering
the free energy $F_{\rm vor}^{(-)}-\Omega L^{(-)}$ in the rotating frame, 
we find that a HQV nucleates
at $\Omega=\Omega_c^{(-)} = \ln R+C_-$, where $\Omega$ is 
scaled by $\hbar/2m R^2$. 

In a SV, both $(+)$
and $(-)$ condensates carry a vortex with a unit winding number
and these components are superposed so that the vortex cores overlap
exactly.
A SV has the free energy
\be
F_{\rm SV} = 4\pi \l[ \l(A_{++}^{(0)}\r)^2 + \l( A_{-+}^{(0)}\r)^2\r]
\l(\ln R + C_S\r).
\ee
The parameter $C_S$ is the singluar vortex core energy and is a function
of $\delta$ and $\hat{h}$. A singular vortex nucleates at a critical
angluar velocity $\Omega_c^S =\ln R + C_S$.

\begin{figure}
\begin{center}
\includegraphics[width=7cm]{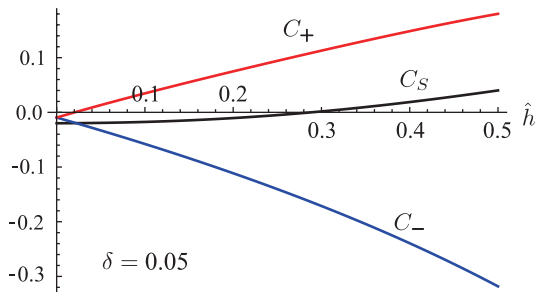}
\end{center}
\caption{(Color online) 
$\hat{h}$-dependence of parameters $C_{+}$ (red), 
$C_-$ (blue) and $C_S$ (black). See Eqs.~(1) and (2)
for definitions of these parameters. The paramagnon
parameter $\delta$ is set to $0.05$.}
\end{figure}
Figure 1 depicts the parameters $C_{\pm}$ and $C_S$ for the paramagnon
parameter $\delta=0.05$, corresponding to low pressure, as functions of
$\hat{h}$.
Observe that when $\hat{h}>0$, there is a range in the diagram
where $C_S > C_-$, which implies that a single HQV, having a phase
factor $e^{i \phi}$ in the $(-)$ condensate, nucleates first
as $\Omega$ is gradually increased from zero.

In this Letter, we consider the case in which the angular velocity
$\Omega$
is further increased to investigate how many HQVs and SVs exist in
the superfluid and the patterns of
the stable configurations of these vortices. 
We again assume
that the $\hat{\vect{l}} = +\hat{\vect{z}}$ at $r \gg 1$.
The free energy of the system with more than one vortex is evaluated
by making use of $C_{\pm}$ and $C_S$ numerically obtained in Fig.~1.
We fix the pressure so that $\delta=0.05$ for
numerical calculations throughout this Letter.
Let $\vect{r}_i$ be the position of the $i$th vortex center.
When the condition $|\vect{r}_i - \vect{r}_j| \gg 1$ is 
satisfied for all pairs $i \neq j$, the London approximation
is valid and energies of the vortices in the $(+)$-condensate
and that of the vortices in the
$(-)$-condensate may be evaluated independently since the coupling
between two condensates appears only through the gradient free energy.
The hydrodynamic
energy associated with the flow around vortices has been
evaluated previously for superfluid $^4$He \cite{hess}. 

Suppose the number of vortices
$n$ satisfies $n \leq 5$. Then the vortices distribute uniformly
on a circle with the radius $r \approx
 \sqrt{(n-1)/2\Omega}\ R$ centered at the origin
of the cylinder, where $\Omega$ is scaled by $\hbar/2mR^2$
as before. Then the free energy of
$n$ SVs in the A$_2$ phase of superfluid $^3$He
in the rotating frame takes the form
\begin{equation}\label{eq:3}
F_n^{(S)}(\Omega, u) = 4\pi \l[\l(A_{++}^{(0)}\r)^2 + \l(A_{-+}^{(0)} \r)^2\r]
F_n(\Omega, u),
\end{equation}
where $u=r/R$ and
\begin{eqnarray}
F_n(\Omega, u) &=& n [ \ln R+C_S + \ln(1-u^{2n})\nonumber\\
& & -(n-1) \ln u - \ln n-\Omega (1-u^2)].
\end{eqnarray}
Here the core energy of the vortices has been taken into account
in the definition of $F_n$.
It has been shown that the function $F_n(\Omega, u)$ has a minimum
at $u$ in the physical region $(0,1)$ when $\Omega$ is greater than some
critical value $\Omega_0(n)$ \cite{hess}. Let 
\begin{eqnarray}
f_n(\Omega) &=& \min_{u \in (0,1) } F_n(\Omega, u)\nonumber\\
&\approx & n (\ln R + C_S -\Omega)   \nonumber\\
& & +\frac{1}{2}n(n-1)[1+\ln(2\Omega) -\ln(n-1)]    \label{eq:approx}
\end{eqnarray}
be the minimum value, where the approximate value
$u \approx \sqrt{(n-1)/2\Omega}$ has been used.
This approximation is verified
numerically to be quite accurate in the given prameter range when $n\geq 2$.
Then the energy of the stable configuration of $n$
singular vortices 
is given by
\be
F_n^{(S)} (\Omega) = 4\pi \l[ \l( A_{++}^{(0)}\r)^2+\l( A_{-+}^{(0)}\r)^2\r]
f_n(\Omega)
\ee
for $n \leq 5$.

Next, let us consider the case in which $n \geq 6$. It was shown for 
superfluid $^4$He that a stable configuration for $6 \leq n \leq 8$
is $n-1$ vortices distributing uniformly in a circle centered at the origin
plus a single vortex at the origin \cite{hess}.
Vortex configuration with less symmetry is expected
as the angular velocity is further increased beyond $n = 8$.
These patterns are verified both experimentally \cite{packard}
and by numerical simulation \cite{cz}. 
When $6 \leq n \leq 8$, $F_n(\Omega, u)$ in Eq.~(\ref{eq:3}) takes the
form \cite{hess}
\begin{eqnarray}
F_n(\Omega, u) &=& n (\ln R+C_S) + (n-1)[\ln(1-u^{2n})\nonumber\\
& & -n \ln u - \ln (n-1) -\Omega (1-u^2)]-\Omega.\nonumber\\
& &
\end{eqnarray}
With the same approximation employed to obtain Eq.~(\ref{eq:approx}),
the free energy mimimizing configulation is given by
$u \approx \sqrt{n/2\Omega}$, which gives the minimum
energy
\begin{eqnarray}
f_n(\Omega) &=& \min_{u \in (0,1)} F_n(\Omega, u)\nonumber\\
&=& n (\ln R +C_S -\Omega)-(n-1)\ln (n-1) \nonumber\\
& & +\frac{1}{2}n(n-1) [1+\ln (2\Omega)-\ln n].
\end{eqnarray}

Similarly the free energies of $n$ HQVs with vortices in $(+)$-condensate and 
$(-)$-condensate are evaluated as
\be\label{eq:hqv}
F_n^{(\pm)}(\Omega) = 4\pi \l(A_{\pm +}^{(0)}\r)^2(f_n(\Omega) + n \Delta
C_{\pm} ),
\ee
where $\Delta C_{\pm} = C_{\pm} -C_S$. Equation (\ref{eq:hqv})
applies to cases in which $2 \leq n \leq 8$.

It is expected that HQVs appear in the vicinity of the parameter 
domain where the free energy difference between the $n$ singluar vortices 
and $n+1$ singular vortices is small. We expect there
are $n$ HQVs with vortices in the $(+)$-condensate and
$n+1$ HQVs with vortices in the $(-)$-condensate
in the process of the
transition from $n$ singular vortices to $n+1$ singular vortices.
We denote this configuration of HQVs as HQV($n+1,n$),
while a configuration with $n$ singular vortices is denoted as SV($n$).
In a sense, HQV($n+1,n$) is roughly regarded as ``SV($n+1/2$)''.
Note that the number of HQVs with vortices in $(-)$ condensate is larger than
that of HQVs with vortices in $(+)$ condensate due to the inequality $C_-<C_+$
(see Fig.~1);
it is energetically favorable to have an extra vortex in the $(-)$ condensate
rather than in the $(+)$ component.

The conditions under which HQV$(n+1,n)$ is stabilized are
\be
F_{n+1}^{(-)}(\Omega) + F_n^{(+)}(\Omega) < F_n^{(S)}(\Omega)
\ee
and
\be
F_{n+1}^{(-)}(\Omega) + F_n^{(+)}(\Omega) < F_{n+1}^{(S)}(\Omega)
\ee
simultaneously. More explicitly, these conditions are written as
\begin{eqnarray}
\lefteqn{\l(A_{-+}^{(0)}\r)^2(f_{n+1}(\Omega)-f_n(\Omega))}\nonumber\\
&+&
\l(A_{++}^{(0)}\r)^2 n \Delta C_+
+ \l(A_{-+}^{(0)} \r)^2
(n+1) \Delta C_- < 0\nonumber\\
& &
\end{eqnarray}
and
\begin{eqnarray}
\lefteqn{
-\l(A_{++}^{(0)}\r)^2(f_{n+1}(\Omega)-f_n(\Omega))}\nonumber\\
& &+
\l(A_{++}^{(0)}\r)^2 n \Delta C_+ + \l(A_{-+}^{(0)} \r)^2
(n+1) \Delta C_- <0,\nonumber\\
& &
\end{eqnarray}
from which the necessary condition for the existence of a
stable HQV($n+1,n$) configuration is
found to be
\be
\Delta F_n \equiv
\l(A_{++}^{(0)} \r)^2 n \Delta C_+ + \l(A_{-+}^{(0)}\r)^2 (n+1) \Delta
C_-<0.
\ee

\begin{figure}
\begin{center}
\includegraphics[width=9cm]{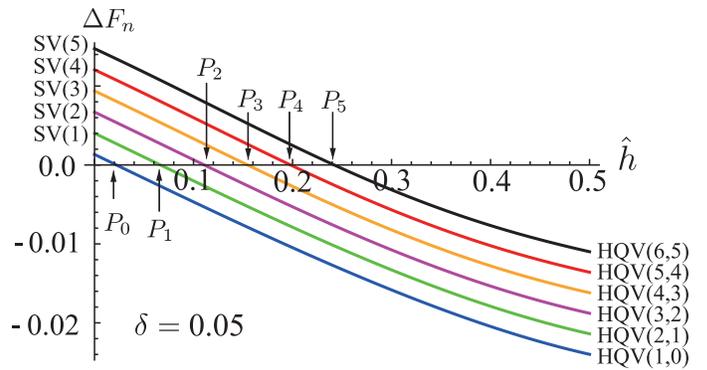}
\end{center}
\caption{(Color online)
$\Delta F_n$ as a function of $\hat{h}$ for $n=0$ (blue), $1$ (green), 
$2$ (purple), $3$ (orange), $4$ (red) and $5$ (black).
Singluar vortices 
are formed when $\Delta F_n >0$ while 
HQV($n+1,n$) is formed when $\Delta F_n < 0$ and $\Omega$
is properly chosen.
The transition point between these two types of vortices are denoted
as $P_n$.
}
\end{figure}
Figure 2 shows the magnetic field ($\hat{h}$)-dependence of $\Delta F_n$
for $n=0,1, \ldots ,5$. 
Note that $\Delta F_n <0$ is a necessary
condition for the existence of HQV($n+1,n$) but not a sufficient
condition. Stability of HQVs and SVs also depends on $\Omega$
as shown in Fig.~3.
It is observed that the stability of HQV lattice is attained
with larger magenetic field as $n$ becomes larger.

\begin{figure}
\begin{center}
\includegraphics[width=7cm]{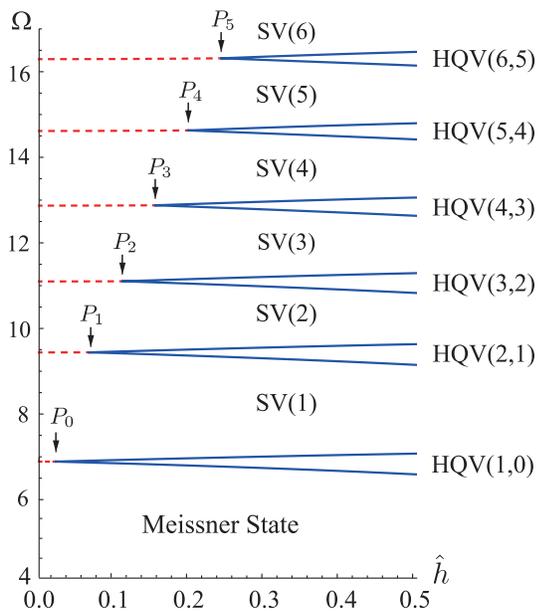}
\end{center}
\caption{(Color online)
Phase diagram of various types of vortices for $\Omega>0$ for
the paramagnon parameter $\delta = 0.05$. 
${\rm{SV}}(n)$ is the domain where $n$ singular vortices are the
most stable configulartion while the wedge shaped domain between
blue solid curves, denoted
${\rm{HQV}}(n+1,n)$, is a region where
a configulatoin of $n+1$ HQVs of $(-)$-vortex and $n$ HQV of $(+)$-vortex are
most stable. A dashed red line is a boundary between two types of SVs
while a solid blue line is a boundary between SV and HQV.
The point $P_i$ denotes the point of the same symbol in Fig. 2.
}
\end{figure}
We plot the phase diagram of various vortex configurations 
in the $\hat{h}$-$\Omega$ plane in Fig.~3. For definiteness,
we have taken $\delta=0.05, R=1000$ and $n=0,1,2,3,4$ and $5$. 
The red dashed line near $P_0$ shows the first critical angluar
velocity $\Omega_c^S$ for formation of a singular vortex, while
the lower blue solid line of the domain HQV($1,0$) is
the first critical angular velocity $\Omega_c^{(-)}$ for formation of
a HQV in the $(-)$ condenstate.

Figure 4 (a) depicts a HQV arrangement for $\hat{h}= 0.5$ and $\Omega = 12.9$, 
for which HQV$(4,3)$ is the most stable configuration. The inner
circle of radius $u_+ = \sqrt{(3-1)/2\Omega}\approx 0.28$ supports 
three $(+)$-HQVs while the outer circle of radius 
$u_- =\sqrt{(4-1)/2\Omega} \approx 0.34$ supports four
$(-)$-HQVs. Here the radius is scaled so that the wall of the cylinder
is at $u=1$.
Figure 4 (b) shows the HQV arrangement
for $\hat{h}=0.5$ and $\Omega = 16.3$, for which HQC($6,5$) is
stablized.
There is a single $(-)$-HQV in the center and five $(-)$-HQVs on the circle
with radius $u_- = \sqrt{6/2\Omega} \approx 0.43$ while
five $(+)$-HQVs distribute on the circle with radius
$u_+ = \sqrt{(5-1)/2\Omega} \approx 0.35$.
\begin{figure}
\begin{center}
\includegraphics[width=6.5cm]{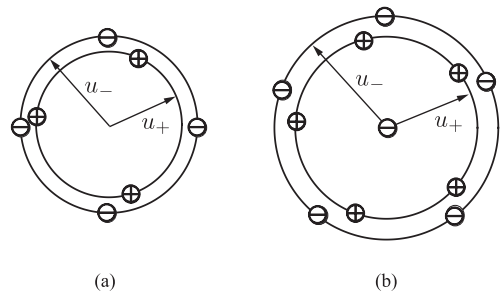}
\end{center}
\caption{(a) Configuration of HQV($4,3$) 
for $\hat{h}=0.5$ and $\Omega = 12.9$. The inner
circle of radius $u_+ \approx 0.28$ supports three $(+)$-HQVs,
denoted by $\oplus$,
while the outer circle of radius $u_-\approx 0.34$ supports
four $(-)$-HQVs, denoted by $\ominus$. 
The relative orientation of $(+)$ HQVs and
$(-)$ HQVs is arbitrary. The wall of the cylinder at $u=1$
is not shown. (b) Configuration of HQV($6,5$) for $\hat{h}=0.5$ and
$\Omega=16.3$. There is a $(-)$-HQV at the center and
five $(-)$-HQVs on a circle with radius $u_- \approx
0.43$ and five $(+)$-HQVs on a circle with radius $u_+
\approx 0.35$.
}
\end{figure}

In summary, we analyzed the stable textures of a thin film
of rotating superfluid $^3$He under magnetic field by
using Ginzurg-Landau free energy.
It was shown that $n$ half-quantum vortices in spin $(+)$ condensate
and $n+1$ half-quantum vortices in spin $(-)$ condensate interpolate
between $n$-signluar votex texture and $n+1$-singluar vortex texture 
as the scaled angluar velocity $\Omega$ is increased 
for sufficiently large scaled magnetic field $\hat{h}$.
The phase diagram for $0 \leq n \leq 5$ has been ploted in
the $\hat{h}$-$\Omega$ plane. We expect these textures may be
experimentally observable by using NMR for example.
Strong magnetic field may be disturbing for NMR measurement.
However turning off or reducing the magnetic field
may not be a serious problem;
transition to different phases of texture, after
the magnetic field is turned off or made small, involves
topology change of the order parameter and hence
unstable texture may persist for considerable length of time \cite{ishi}.
Dynamics of the texture after the magnetic field is reduced
is beyond the scope of the present Letter and will be studied
in a separate publication. Analysis of the oscillation modes
of the vortices on a circle is also an interesting problem,
which might have some observable consequence \cite{tk}.

We would like to thank Kenichi Kasamatsu for useful 
discussions. M.~N. is grateful to Osamu Ishikawa and Kazumasa
Miyake for enlightening discussions.
This work is supported by the ``Topological Quantum Phenomena'' (No. 22103003) Grant-in Aid for Scientific Research on Innovative Areas from the 
Ministry of Education, Culture, Sports, Science and Technology (MEXT) of Japan.


\begin{thebibliography}{99}

\bibitem{vw} D. Vollhardt and P. W\"olfle, {\it The Superfluid
Phases of Helium 3} (Taylor \& Francis, London, 1990).

\bibitem{vol1} G. E. Volovik, {\it Exotic Properties of Superfluid
$^3{\rm He}$} (World Scientific, Singapore, 1992).

\bibitem{vol2} G. E. Volovik, {\it The Universe in a Helium Droplet}
(Oxford University Press, Oxford, 2003). 

\bibitem{vm}
G. E. Volovik and V. P. Mineev, JETP Lett. {\bf 24}, 561 (1976).

\bibitem{sv} M. M. Salomaa and G. E. Volovik, Phys. Rev. Lett.
{\bf 55}, 1184 (1985).


\bibitem{bec2} K. Kasamatsu, M. Tsubota, and M. Ueda, Int. J. Mod.
Phys. B {\bf 19}, 1835 (2005).

\bibitem{bec3} M. Eto, K. Kasamatsu, M. Nitta, H. Takeuchi, and M. Tsubota,
Phys. Rev. A {\bf 83}, 063603 (2011).

\bibitem{bec4} M. Cipriani and M. Nitta, arXiv:1303.2592.


\bibitem{sc1} S. B. Chung, H. Bluhm, and E.-A. Kim,
Phys. Rev. Lett. {\bf 99}, 197002 (2007).

\bibitem{sc2} S. B. Chung and S. A. Kivelson,
Phys. Rev. B {\bf 82}, 214512 (2010).

\bibitem{sc3} J. Jang, D. G. Ferguson, V. Vakaryuk, R. Budakian, S. B. Chung,
P. M. Goldbart, and Y. Maeno, Science {\bf 331}, 186 (2011).


\bibitem{th1} M. C. Cross and W. F. Brinkman, J. Low Temp. Phys. {\bf 27}, 683
 (1977).

\bibitem{th2} M. M. Salomaa and G. E. Volovik, Phys. Rev. Lett. {\bf 55}, 1184
(1985).

\bibitem{th3} V. Vakaryuk and A. J. Leggett, Phys. Rev. Lett. {\bf 103}, 057003
(2009).

\bibitem{ex1}
P. J. Hakonen, K. K. Nummila, J. T. Simola, L. Skrbek, and G.
Mamniashvili, Phys. Rev. Lett. {\bf 58}, 678 (1987).

\bibitem{ex2}
R. Ishiguro, O. Ishikawa, M. Yamashita, Y. Sasaki, K. Fukuda, M.
Kubota, H. Ishimoto, R. E. Packard, T. Takagi, T. Ohmi, and T.
Mizusaki, Phys. Rev. Lett. {\bf 93}, 125301 (2004).

\bibitem{on} K. Kondo, T. Ohmi, M. Nakahara, T. Kawakami,
Y. Tsutsumi, and K. Machida,
J. Phys. Soc. Jpn. {\bf 81}, 104603 (2012).

\bibitem{hess} G. B. Hess, Phys. Rev. {\bf 161}, 189 (1967).

\bibitem{packard} R. E. Packard, Physica {\bf 109} \& {\bf 110},
1474 (1982).

\bibitem{cz} L. J. Campbell and R. M. Ziff, Phys. Rev. B {\bf 20},
1886 (1979).


\bibitem{ishi} O. Ishikawa (private communication).


\bibitem{tk} C. D. Andereck, J. Chalupa, and W. I. Glaberson, Phys. Rev. Lett., {\bf 44}, 33 (1980).


\end{thebibliography}
\end{document}